\documentclass[12pt]{iopart}
\usepackage{lineno}
\usepackage{graphicx}
\usepackage{hyperref}

 \expandafter\let\csname equation*\endcsname\relax
  \expandafter\let\csname endequation*\endcsname\relax
\usepackage{amsmath}

\newcommand{\m}{m$^{-3}$ }

\begin{document}

\title{Fundamental Photoemission Brightness Limit from Disorder Induced Heating }
\date{\today}

\author{J M Maxson$^1$, I V Bazarov$^{1}$,
W Wan$^2$,   H A Padmore$^2$ and C E Coleman-Smith$^3$}

\address{$^1$ Cornell Laboratory for Accelerator-Based Sciences and Education,
  Cornell University, Ithaca, New York 14853}
\address{$^2$ Advanced Light Source, Lawrence Berkeley National Lab, 1 Cyclotron Road MS 2R0400, Berkeley, CA 94720}
\address{$^3$ Department of Physics, Duke University,  Durham, NC 27708}
\ead{jmm586@cornell.edu}


%




\begin{abstract}
We determine the limit of the lowest achievable photoemitted electron temperature, and therefore the maximum achievable electron brightness, due to heating just after emission into vacuum, applicable to dense relativistic or nonrelativistic photoelectron beams. This heating is due to poorly screened Coulomb interactions equivalent to disorder induced heating seen in ultracold neutral plasmas. We first show that traditional analytic methods of Coulomb collisions fail for the calculation of this strongly coupled heating. Instead, we employ an N-body tree algorithm to compute the universal scaling of the disorder induced heating in fully contained bunches, and show it to agree well with a simple model utilizing the tabulated correlated energy of one component plasmas. We also present simulations for beams undergoing Coulomb explosion at the photocathode, and demonstrate that both the temperature growth and subsequent cooling must be characterized by correlated effects, as well as correlation-frozen dynamics. In either case, the induced temperature is found to be of several meV for typical photoinjector beam densities, a significant fraction of the intrinsic beam temperature of the coldest semiconductor photocathodes. Thus, we expect disorder induced heating to become a major limiting factor in the next generation of photoemission sources delivering dense bunches and employing ultra-cold photoemitters.
\end{abstract}
\maketitle

\section{Introduction}
Beam brightness is a principle figure of merit for relativistic photoelectron sources for use in high brilliance linear accelerators or for ultrafast electron diffraction (UED) experiments. It is qualitatively defined as the average particle flux per phase space volume. For high brilliance linear accelerators used for x-ray production, the brightness of the x-ray beam is directly determined by that of the electron beam. For UED setups, the brightness of the electron beam is main beam parameter that determines the visibility of diffraction pattern per electron pulse.    

Such electron sources are comprised of a photoemitting material placed in an accelerating gradient, where both direct current (DC) and radio frequency (RF) accelerating fields are used for various applications. For either x-ray or electron diffraction experiments, it is often the 4 dimensional transverse normalized brightness that is most pertinent. For a given beam current $I$, we can define the ``micro-brightness" as the phase space density itself, $\rho=\frac{d I}{d V_4}$, where $dV_4=dxdydp_xdp_y$ is the phase space volume element. The normalized total beam brightness can then be defined as a statistical average of the micro-brightness:
\begin{equation}
\mathcal{B}_{n,4D}=\frac{I^{-1}}{\left(mc\right)^2}\int \rho(x, y, p_x, p_y)^2 dV_4
\end{equation}
For a bunched beam with some bunch repetition rate $f$, the average current can be written as $I_{av}=qf$, where $q$ is the charge of the bunch. In a previous work \cite{ivan}, it was shown that the maximum achievable beam brightness (either microbrightness or total) can be written:
\begin{equation}
 \mathcal{B}_{\text{n,4D }} \bigg|_{\text{max}}=\frac{mc^2 f\epsilon_0 E_{\text{acc}}}{2\pi kT} 
\label{maxB}
\end{equation}
where $E_{\text{acc}}$ is the accelerating electric field directly at the photocathode, which sets the maximum supportable charge density at the photocathode. The minimum divergence is set by $kT$, the temperature of photoemitted electrons, an intrinsic property of the choice of photocathode and laser wavelength. A nonzero temperature arises fundamentally from the electron momentum spread inside the photoemitting material, and can then be significantly increased due to excess laser energy above the photoemission threshold, as well as electron scattering off of imperfections in the emitter. 

As it is one of the two independent parameters of the maximum achievable beam brightness per bunch, the photoemission temperature has been the focus of much work in photoelectron sources. Great progress has been made by those working with negative electron affinity semiconductor photocathodes and those pursuing electron emission from laser cooled atoms. Several semiconductor photocathode groups have measured photoemission of equivalent temperatures well below $kT=25$ meV, or the thermal energy at room temperature, either via the cryogenic cooling of the photocathode \cite{cryo}, or via the maintenance of a pristine photoemissive conditions in ultra high vacuum, allowing the low effective mass of conduction electrons to produce an effect equivalent to Snell's law in which the velocity spread of electrons is drastically reduced at the vacuum interface \cite{narrowcone}. Furthermore, using laser cooled atoms which are photoionized, there has been production of electrons at temperatures equivalent to $\sim1$ meV \cite{MOT}. 

The number of electrons per bunch varies widely across various applications, with a range approximately between $10^6\rightarrow10^9$ electrons, which for moderate  to high flux applications often corresponds to densities in the range of $n_0 =10^{17}\rightarrow10^{20}$ m$^{-3}$. For a near-zero temperature bunch with such high density, we expect some contribution of individual stochastic Coulomb interactions from close encounters just after emission into vacuum to add to the total effective photoemission temperature. In this work, we determine this amount of stochastic heating as a function of beam density and initial temperature, as well as the nature of its evolution in time.

We expect this effect to be most prevalent when the electrostatic potential energy of neighboring particles is comparable to their thermal energy, that is $kT\sim e^2/4\pi \epsilon_0 a$, where $a$ is the Wigner-Seitz radius, $a=(3/4 \pi n_0)^{1/3}$. Thus for a given density, we expect the heating to be of the order $e^2/4\pi \epsilon_0 a$, and should thus scale with the cubic root of the density, and should be independent of the number of particles in the bunch. For a rough estimate of the importance of the effect, the plasma coupling parameter $\Gamma$ can be used, defined as the ratio of $kT$ to the average pair interaction potential. It ranges from  $\Gamma=e^2/4 \pi \epsilon_0 a k T=0.2\rightarrow 2$ given an electron temperature of 5 meV. Thus, for applications requiring a large charge density, we expect this stochastic heating to serve as a hard limit to the lowest attainable electron temperature, and limiting the maximum attainable beam brightness. 

\section{Failure of Analytic models for Coulomb Collisions}
We will now describe how traditional collisional methods in beam physics, familiar to many accelerator practitioners, fail for the case of a cold dense beam. Readers familiar  with the inability of such methods to accurately describe our strongly correlated case can bypass this section.  A simple model for a bunch that has just been photoemitted into vacuum is a static, uniform, randomly distributed electron sphere with very small initial temperature $kT\sim 0$, in a constant accelerating field. We first draw a sharp distinction between the stochastic heating in question and the effects of space charge, which is the collective, mean field effect of Coulomb repulsion. The space charge approximation, applied in most beam physics calculations, self-consistently calculates the interparticle interaction based on the local single particle beam density, $\nabla^2 \Phi(\mathbf r)=en(r)/\epsilon_0$, where $\Phi$ is the total electrostatic potential of a particle at $\mathbf r$. This approximation requires that the individual electron interaction is heavily screened, meaning that the Debye screening length $\lambda=\sqrt{\epsilon_0 k T/n_0 e^2}$ is much larger than the interparticle separation. However, for the lowest temperatures attained in semiconductor photoemission $kT\sim5$ meV \cite{narrowcone}, and the lowest of the above densities $n_0=10^{17}$ m$^{-3}$, the Debye length is already on the order of the interparticle separation. In this case, the collective field will describe the overall density evolution, but cannot capture the growth of the stochastic component of the velocity. 

The effect of Coulomb interactions in particle beams has been treated analytically in various schemes. Perhaps the most famous is the diffusive Fokker-Planck method. The Fokker-Planck method assumes that the effects of Coulomb collisions can be treated via the calculation of effective velocity diffusion and dynamical friction terms. This approach requires that the shifts in velocity due to Coulomb collisions are small compared to the overall velocity spread of the beam. This assumption is maximally violated for the case of a cold dense beam as described above, in which transverse velocities begin near zero. 

Non diffusive, two-particle methods have also been developed that do not require the assumption of Debye screening. A clear presentation of these is given in \cite{jansen}. The method applicable over the largest parameter space is the Extended Two-Particle Approximation (ETPA), developed by Jansen.  This method relies on the calculation of the displacement (either velocity displacement, or position displacement) of a test particle in the presence of a single field particle over some collision time $\delta t_c$. If we calculate the velocity displacement $\Delta v$ of a particle pair initially at rest, the ETPA allows the formation of the probability distribution $\rho(\Delta v)$, by averaging $\Delta v$ over all possible separations and initial velocities in the beam. Each of these encounters is assumed to be statistically and dynamically independent. The second moment of the velocity distribution $\langle \Delta v^2 \rangle$ is then a measure of the temperature. We will proceed with a sketch of this calculation to highlight the failure of some of its assumptions in the cold dense beam case.

For two particles initially at rest (i.e. $kT\sim 0$) with initial separation $r_i$, using the dimensionless variables $\tilde{r}= r/r_i$ and $\tilde{t}=t \left(  e r_i^{-3/2}/\sqrt{2\pi \epsilon_0 m}\right) $ the equation of motion for their separation is given by:
\begin{equation}
1=\left(\frac{d\tilde{r}}{d\tilde{t}}\right)^2+\frac{1}{\tilde{r}}
\end{equation} 
This equation is integrable for the function $\tilde{t}(\tilde{r})$, which is not analytically invertible, but is trivial to invert numerically to obtain $\tilde{r}(\tilde{t})$. With a global choice of $\delta t_c$, and with a test particle chosen at the origin, we can obtain $\Delta v$ as a function of $r_i$ by replacing the scaling factors.  Then, we average over the entire distribution of $r_i$:
\begin{equation}
 \langle  \Delta v ^2 \rangle = \int n(\mathbf r)\Delta v(r, \delta t_c)^2 d^3\mathbf r   =\frac{2 e^2 n_0}{\epsilon_0 m} \int  r^2dr \left( \frac{d\tilde{r}(\tilde{t})}{d\tilde{t}}\frac{1}{\sqrt{r}} \right)^2  
\label{ETPA}
\end{equation}
where $\tilde{t}$ is also evaluated at $\delta t_c$ and $r_i$. The velocity kick $\Delta v$ falls off sufficiently fast with large separation, and the integral measure $r^2dr$ ensures that the averaging does not diverge at small r, and thus we can integrate over all space. The equation \ref{ETPA} is the statistical average of all two particle interactions in a beam for during some time $\delta t_c$.

\subsection{Unbound Trajectories}

The expression given in \ref{ETPA} and similar uncorrelated two-particle collision methods fail in the cold dense beam case for two reasons. First, it assumes that each pair-wise interaction is statistically independent from each other. There can be no dynamic correlation between separate pair interactions. However in the cold dense beam case a large contribution to the final temperature can be given by simultaneous 3-body (or higher) interactions. The assumption of statistical and dynamical independence allows for the unbound expansion particles with very small initial separation, which are the most pertinent interactions in the cold dense beam case. 

Even from the definition of the scaled coordinates, it is clear that any choice of $\delta t_c$ corresponds to a some electron separation $r_c$ below which all collisions taking place in that time will be sufficiently complete collisions, in which all potential energy is converted to kinetic energy. Alternatively put, given $\delta t_c$ there will always be some electron seperation small enough to make $\tau$ arbitrarily large. If we assume the initial distribution of electron separations to be uniform over all length scales, there is no unambiguous choice of cutoff in for $r_c$ to avoid such unphysical free expansion, and thus no inherent timescale for two particle collisions across the entire bunch. For perspective, in other Coulomb collision calculations, this ambiguity is seen the calculation of the so-called ``Coulomb Logarithm", defined as $\ln \left(b_{max}/b_{min}\right)$, or the logarithm of the ratio of the maximum to minimum impact parameters over the whole bunch. However, as it appears under the logarithm in such problems, the minimum distance is often not considered to be a sensitive parameter \cite{reiser}.

In a similar method to the ETPA, in \cite{massey} Massey et al. calculate the uncorrelated root mean square fluctuation in the interaction force, and from it they obtain a stochastic energy spread. The authors directly impose a minimum interaction distance which corresponds to those collisions for which half of the potential energy is converted to kinertic energy over a certain time. Collisions with more of a fraction of potential energy release (i.e. smaller sepration) are ignored. They readily acknowledge the ambiguity of this choice, and argue that the effect is small for the beams in their study. However, for a cold beam just after photoemission, it is this fast-timescale release of potential energy as close neighbors rearrange that we are interested in calculating. Thus, an average over independent two particle interactions is not sufficient here.

\subsection{Scaling with Density}

Furthermore, even if one does make a choice of $\delta t_c$ based on some other reasoning, the fact that this scheme involves taking a moment of the single particle distribution means one will always find $\langle v^2 \rangle \sim n_0$, whereas the effect we desire to calculate should have $\langle v^2 \rangle \sim n_0^{1/3}$, argued above. This scaling of the velocity spread with density occurs in the Fokker-Planck method (as shown in \cite{reiser} in Eq. 5.243), in the ETPA (as shown in \cite{jansen} in Eq. 5.5.11), and in the work by Massey (reference \cite{massey} in Eq. 19).

A method that produces the correct scaling with density, but is also flawed, is again given by Jansen in \cite{jansen}. In what he calls the ``thermodynamic limit", Jansen takes the difference of of an the total electrostatic potential enery of an initially uniform distribution of charge with no screening, and a final state of a Debye screened distribution with some $kT$, and sets this difference equal to the the total thermal energy, $\frac{3}{2}NkT$. However, instead of using a single particle density, Jansen implicitly uses the two-particle correlated density for a Debye screening:

\begin{equation}
\frac 3 2 nV k T=U_{\text{i}}-U_{\text{f}}=\frac 1 2 n^2 V \int^\infty _0 d^3 r \frac{e^2}{4 \pi \epsilon_0 r}\left(1- \exp[-\phi(r)/kT] \right) \label{jan1} 
\end{equation}
Here $U_i$ and $U_f$ stand for the initial and final potential energy in the system, respectively given by the first and second terms of the integral. $\phi(r)$ is the interaction potential of two electrons seperated by $r$, and is assumed to have the form: $\phi(r) =e^2\exp(-r/\lambda)/4\pi \epsilon_0 r$. The factor $n\exp(-\phi(r)/kT)$ is the final two particle correlated density. This equation has a solution of the form:
\begin{equation}
kT = \frac{e^2}{4 \pi \epsilon_0} \left(4 \pi \alpha^2\right)^{1/3} n^{1/3}
\end{equation}
Where $\alpha$ is a dimensionless number determined by numerical solution of the above, $\alpha=0.08702$ \footnote{An incorrect value of alpha is quoted for $\alpha$ in \cite{jansen},  which explains the use of the Debye relations, though the final value of $\lambda$ shows that they are not applicable.} This number corresponds to an coupling factor at thermodynamic equilibrium of $\Gamma_{eq}=3.53$. However, if one evaluates the number of particles in the Debye sphere, one finds $N_D=\frac{4}{3}\pi \lambda^3 n_0 \approx 0.03$, whereas the Debye approximation requires that the number of particles in the Debye sphere must be large. Thus, we may not apply the Debye/Yukawa form for $\phi(r)$ nor for the two particle density function used in \ref{jan1}. It is the use of a two particle correlated density function that provides the correct scaling with density, however, it is difficult to analytically compute the correlation in general. It must be found by some other numerical means \cite{teller}.

\section{Disorder Induced Heating}

The heating associated with the relaxation of a random, near-zero temperature  distribution of charges is well known to the ultracold neutral plasma (UNP) community. In such systems, a cold gas is laser ionized, and after a time on the order of the $\tau=2\pi\omega_p^{-1}=2 \pi \left(n_0 e^2/ m_e \epsilon_0 \right)^{-1/2}$. In the traditional plasma physics terminology, this effect is referred to as disorder induced heating (DIH), and the effect is seen for $\Gamma$ of order unity or larger \cite{referencefarm}, and we will argue that an exact analog of this effect is present in practical photoemisison.

Disorder induced heating has been experimentally observed for the ions in a neutral plasma \cite{verifymurillo}. In cold neutral plasmas, the electrons equilibrate much faster than the ionic $\omega_p^{-1}$, and then serve to screen ion-ion interaction. An expression for the amount of DIH in ions was first given for Yukawa systems (for which electron screening, but not electron-ion recombination, is considered) in \cite{Murillo}. The initial ionic distribution is uncorrelated, and can be defined as the zero energy state. As the ions relax, the ion distribution begins to develop order, and the resulting correlations have an associated binding energy. This correlation binding energy can be calculated from the two particle density function $g(r)$, and is only a function of $\Gamma$ and the electron screening parameter $\kappa=\lambda/a$:
\begin{equation}
\frac{U_{c}}{NkT} \equiv \bar U= \frac{1}{2}\frac{e^2 n_0}{4\pi \epsilon_0 }  \int \frac{ g(r, \Gamma, \kappa) d^3r}{r}
\end{equation}
This binding energy is balanced by an increasing ion temperature.  Calculating the increase in temperature thus only requires knowledge of $\bar U$ for a given electron screening. Analytic calculation of $g(r)$ for strongly coupled systems is difficult, and alternatively, $\bar U$ has been tabulated via molecular dynamics (MD) simulations \cite{Murillo}. 

It has been demonstrated that a trapped, charged plasma, such as an electron bunch, is equivalent to the one-component plasma (OCP) model, in which the charges exist in a uniform neutralizing background \cite{review}. The external containing potential allows the initial uncorrelated state to be viewed as zero total energy. As electrons relax, correlations develop, and the presence of a ``Coulomb hole" in $g(r)$ for $r<a$ creates an effective correlation binding energy equal to that of a one component plasma.  In the case of an OCP, there is no second species to provide additional screening, and thus $\bar U$ is only a function of $\Gamma$. Owing to this simplification, MD data has been fit to a power series relation for $\bar U=a\Gamma+b\Gamma^{1/3}+c$, where the coefficients are given in \cite{review},         $a=-0.90$, $b=0.590673$, $c=-0.26569$. We may now write down the expression for the the final coupling $\Gamma_f$, and thus the final equilibrium temperature, for a fully confined, initially uncorrelated ($\bar U_i =0$), uniform distribution of charges with initial coupling $\Gamma_i$, analogous to that presented in \cite{Murillo}: 

\begin{equation}
\bar U_f =\frac{3}{2} \left(\frac{\Gamma_f}{\Gamma_i}-1  \right)= a\Gamma_f+b\Gamma_f^{1/3}+c
\label{mymurillo}
\end{equation}

\begin{figure}[ht]
\includegraphics[width=0.8\linewidth, clip]{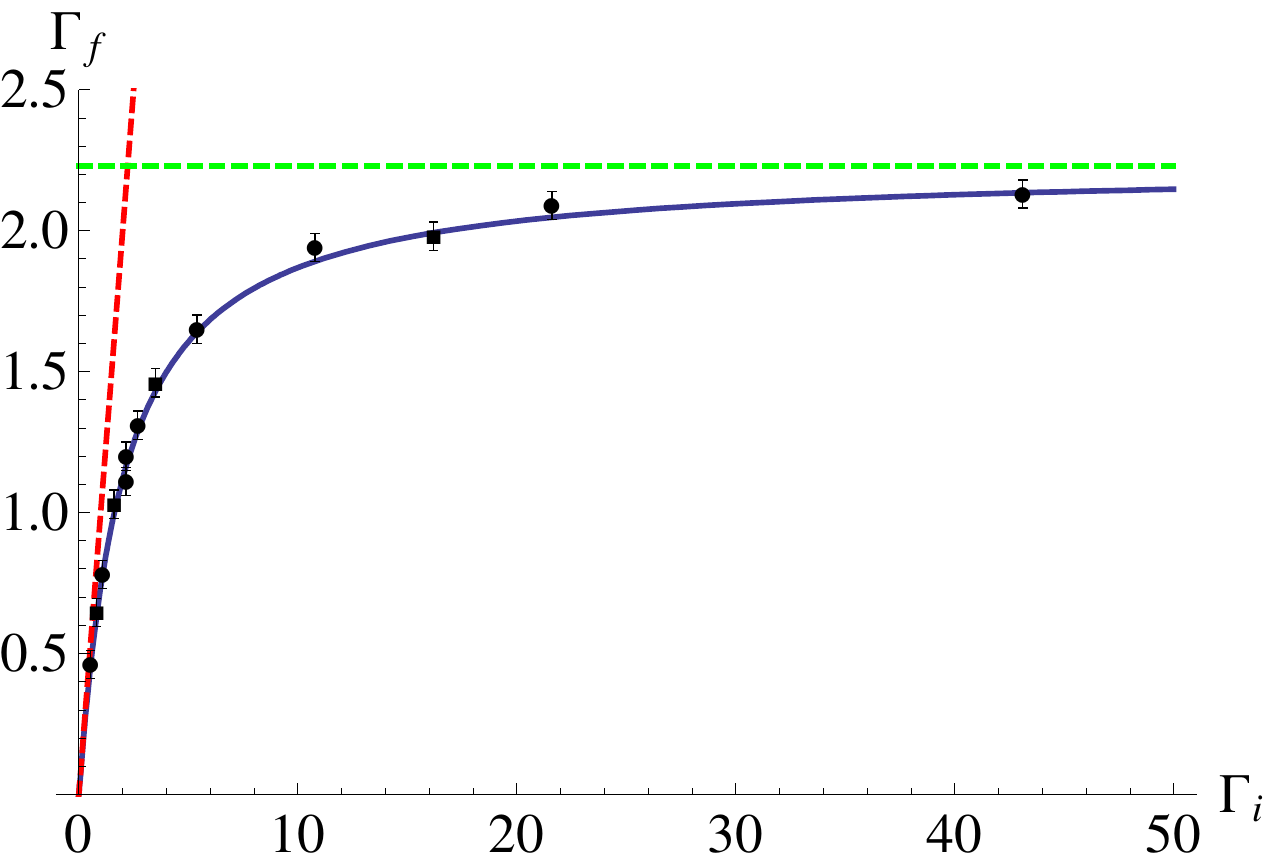}
\caption{\label{gamma} Final coupling $\Gamma_f$, vs initial coupling $\Gamma_i$, given by  \ref{mymurillo} (blue line), with $\Gamma_f=\Gamma_i$ for $\Gamma_i\rightarrow 0$ (red, dotted), and $\Gamma_f=2.23$ for $\Gamma_i\rightarrow \infty$. Dots are simulation results of a fully contained, equipartitioned electron sphere, density $n_0=10^{20}$ m$^{-3}$, and initial temperatures between kT=0.25 meV and 20meV. Squares are simulation results for multiple densities, where $kT_z=0$, and $kT_x=kT_y$, and where $\Gamma$ is calculated from the average of the three directions. Error bars are an estimate of the uncertainty in final temperature determination due to the residual oscillations, as in figure \ref{gammainf}. }
\end{figure}

It is important to note the power series relation is quoted only for $\Gamma>1$, however we find that for $\Gamma<1$, the above expression is well approximated by $\Gamma_i=\Gamma_f$, or the limit of no DIH, as we expect. The final coupling given in \ref{mymurillo} is plotted in figure \ref{gamma}. For an ideal bunch with zero initial temperature, $\Gamma_i\rightarrow \infty$,  \ref{mymurillo} can be solved to give $\Gamma_f\approx2.23$.  

Practical relativistic photoelectron sources contain a number of complicating factors.
Inside the emitting material (either a cold gas or crystal), the electron coupling, and thus DIH, will be dramatically reduced by the presence of the ionic/nuclear potential, and thus here we only need to consider DIH developing in vacuum.  The beam density during emission will vary in time and with position across the bunch. The temperature associated with DIH should be reached on the order of $\omega_p^{-1}$ after electron emission into vacuum, which for typical beam densities is on the picosecond scale. On this scale, we can neglect relativistic effects, such as the $1/\gamma^2$ damping of the interaction force \cite{reiser}. The coupling given by $\Gamma_f=2.23\equiv \Gamma_{eq}$, for a given local beam density, can then be viewed as the fundamental photoemission temperature limit. We consider the effects of a changing local beam density, as well as the nonrelativistic effects of acceleration  during the evolution of DIH  below, and we find it to correspond well with the prediction of \ref{mymurillo}. Rewriting it in terms of practical units, the heating induced in a zero temperature bunch is given by $kT[\text{eV}]=1.04 \times10^{-9}  (n_0 [ \text{m}^{-3} ] )^{1/3}$. Along with  \ref{maxB}, this forms the fundamental phoemission brightness limit. 

To test the prediction of \ref{mymurillo}, we have chosen to use a tree-algorithm electrostatic simulation package, with nonrelativistic, spherical, uniform particle bunches. Traditional particle-in-cell integrators have an intrinsic length scale---the spatial grid on which local fields are calculated. However, tree algorithms lack a spatial grid, and thus are better suited to our initial conditions, where both close interactions and long range correlations are significant. Several plasma treecodes have been developed, see for instance \cite{plasmatreecode}, and one has been included in a relativistic particle accelerator code \cite{gpt}. However, considering the simplicity of the problem, we elected to directly modify a code originally written for gravitational interaction \cite{barnesonly}. 

We consider a randomly distributed spherical bunch of $10^5$ particles with uniform density, and vary only the density ($10^{17}\rightarrow 10^{20}$ m$^{-3}$) and initial Gaussian velocity distribution ($kT\in \{0.25,   20\}$ meV or $kT=0$) , where we use the definition $kT=m_e\left(\langle v_i ^2\rangle-\langle x_i v_i \rangle ^2/\langle x_i^2 \rangle  \right)$, where $m_e$ is the electron mass, and $i$ is a Cartesian coordinate. There is no breaking of spherical symmetry in our simulations, thus equipartition will always hold. For a fully contained uniform bunch, $\langle x_i v_i \rangle\approx 0$, and the temperature is just a measure of average electron kinetic energy, with no spatial dependence. Containment is done in simulation via the application of a radial external force equal and opposite to the SC field. The parameters of the interparticle force calculation are the Barnes-Hut opening angle $\theta$, and the leap frog integrator time step $dt$. The opening angle is the force calculation accuracy parameter that determines whether an electron is treated individually or lumped together with other similarly distant particles \cite{barneshut}. No effect was seen from a force-softening distance $p$, such that the force between two particles is $f_{12}\propto (r_{12}+p)^{-2}$, from $p=p=a/10^{6}$ up to $p=a/100$.  For all simulations presented, we have demonstrated convergence using $dt=\tau/120$, $\theta=23^{\circ}$, $N=10^5$.

\begin{figure}[ht]
\includegraphics[width=0.80\linewidth, clip]{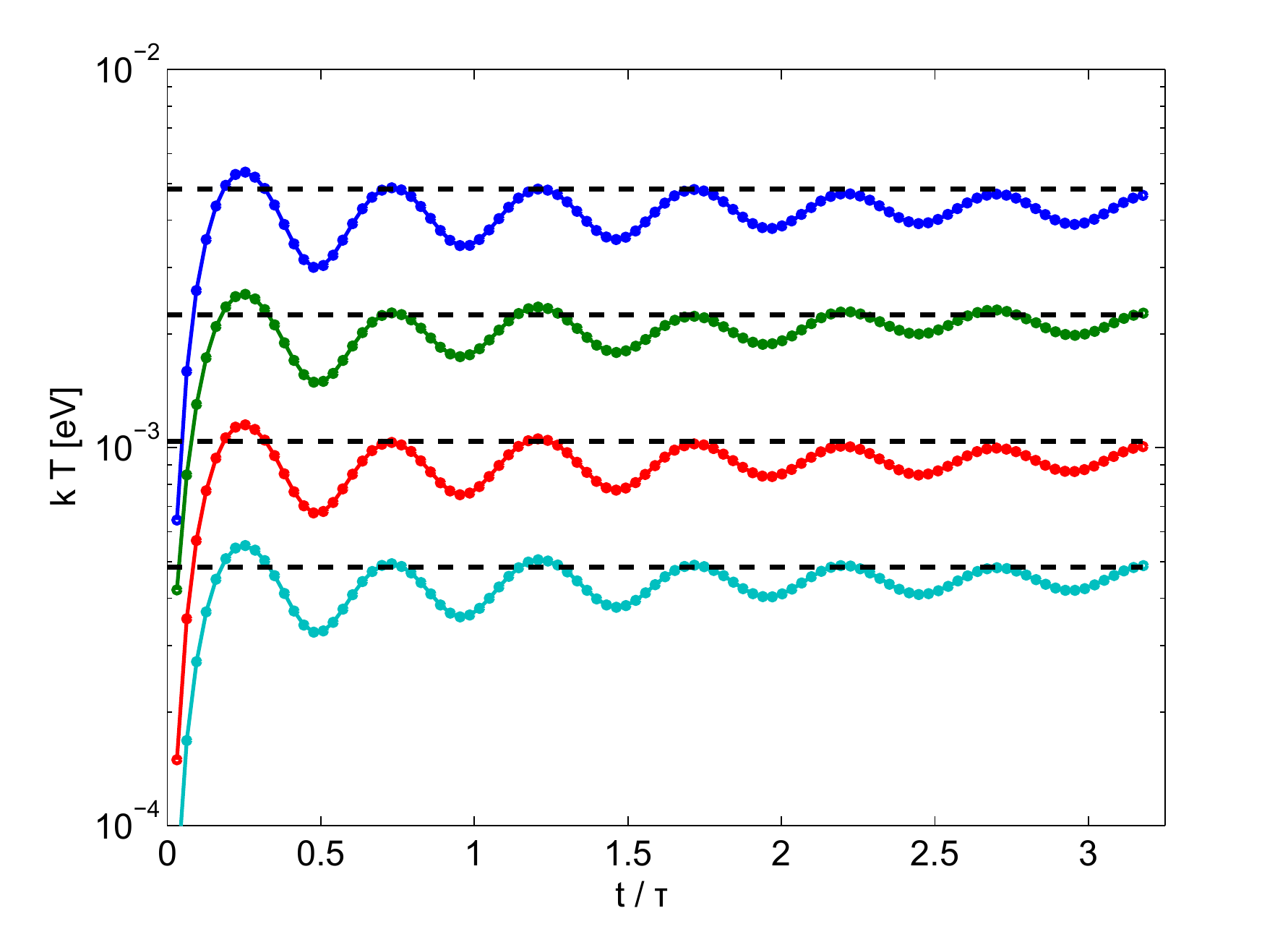}
\caption{\label{gammainf} Temperature vs. number of plasma periods from treecode simulation using $10^{5}$ contained electrons, and $n_0=10^{20}$ (top curve),  $10^{19}$, $10^{18}$, $10^{17}$   m$^{-3}$ (bottom curve), with zero initial temperature, $\Gamma_i\rightarrow \infty$. The equilibrium value predicted by  \ref{mymurillo}, $\Gamma_f\approx2.23$ is shown (dotted lines). }
\end{figure}

%

For an initial distribution with zero temperature, the final temperature for several applicable beam densities is plotted in figure \ref{gammainf}.  We note that for extremely dense beams,  $n_0>10^{19}$ $m^{-3}$, the DIH is comparable to the temperature of photoemission from the coldest semiconductor cathodes $kT\sim5$ meV. Furthermore, we note the oscillation of the electron temperature at $2\omega_p$, which has been measured experimentally in UNP systems \cite{verifymurillo}. These oscillations last far longer than those seen in UNPs, as the bunch remains uniform in simulation (rather than Gaussian), and there is thus no spatial temperature smearing. 

 
For bunches with nonzero initial temperature, the amount of additional heating can be calculated in the final coupling, plotted along with the prediction of \ref{mymurillo} in figure \ref{gamma}. All simulations here had a constant density $n_0=10^{20}$ \m, with varied initial Maxwellian velocity distribution. The duration of the temperature oscillations visible in figure \ref{gammainf} is reduced with increasing initial temperature, though the exact dependence was not extracted in this study. In practical systems, initial acceleration in a voltage gap $V$ will have cooled any photocathode emission temperature along the acceleration direction by a factor $kT_i/eV$, making $kT_{||}\sim0$ almost instantaneously in high gradient electron sources, whereas the transverse velocity spread will remain unaffected \cite{reiser}.   Simulations were performed for contained bunches of multiple densities with $kT_x=kT_y \neq 0$ and $kT_z=0$. Equivalent dynamics to figure \ref{gammainf}, as well as agreement with \ref{mymurillo} were found if the average of the temperatures in all three directions is used to calculate $\Gamma_i$ and $\Gamma_f$, as shown in Fig \ref{gamma}. The timescale of equipartition here also depends on $kT_i$ and $n_0$, but for the densities considered above and initial temperatures comparable to DIH heating, it was seen that equipartition occurs between $\tau$ and $3\tau$. A bunch instantaneously emitted into a uniform accelerating field will not have the DIH evolution altered, a fact that was also verified in simulation.

Real electron bunches just after emission are very infrequently fully transversely  contained in the sense described above, and are not emitted instantaneously. The overall space charge force can double the beam radius on the timescale of $\omega_p^{-1}$, and acceleration can significantly lengthen the pulse afterwards. Thus, we must consider the process of DIH in bunches with time dependent density reduction. A system of fully coupled charges will have an equilibrium temperature that decreases as $kT\propto 1/\Gamma_{eq} a$, as given by above. However, it is well known that system of noninteracting charges expanding under linear forces will have an equilibrium temperature that decreases as $kT\propto 1/a^2$ \cite{reiser}, as a consequence of RMS emittance conservation, which therefore leaves the phase space density or brightness unchanged.  During the explosion, the beam is in a highly nonequilibrium state, but our definition of temperature above still applies, where here $\langle r v_r \rangle \neq 0$.  

\begin{figure}[ht]
\includegraphics[width=1.0\linewidth, clip]{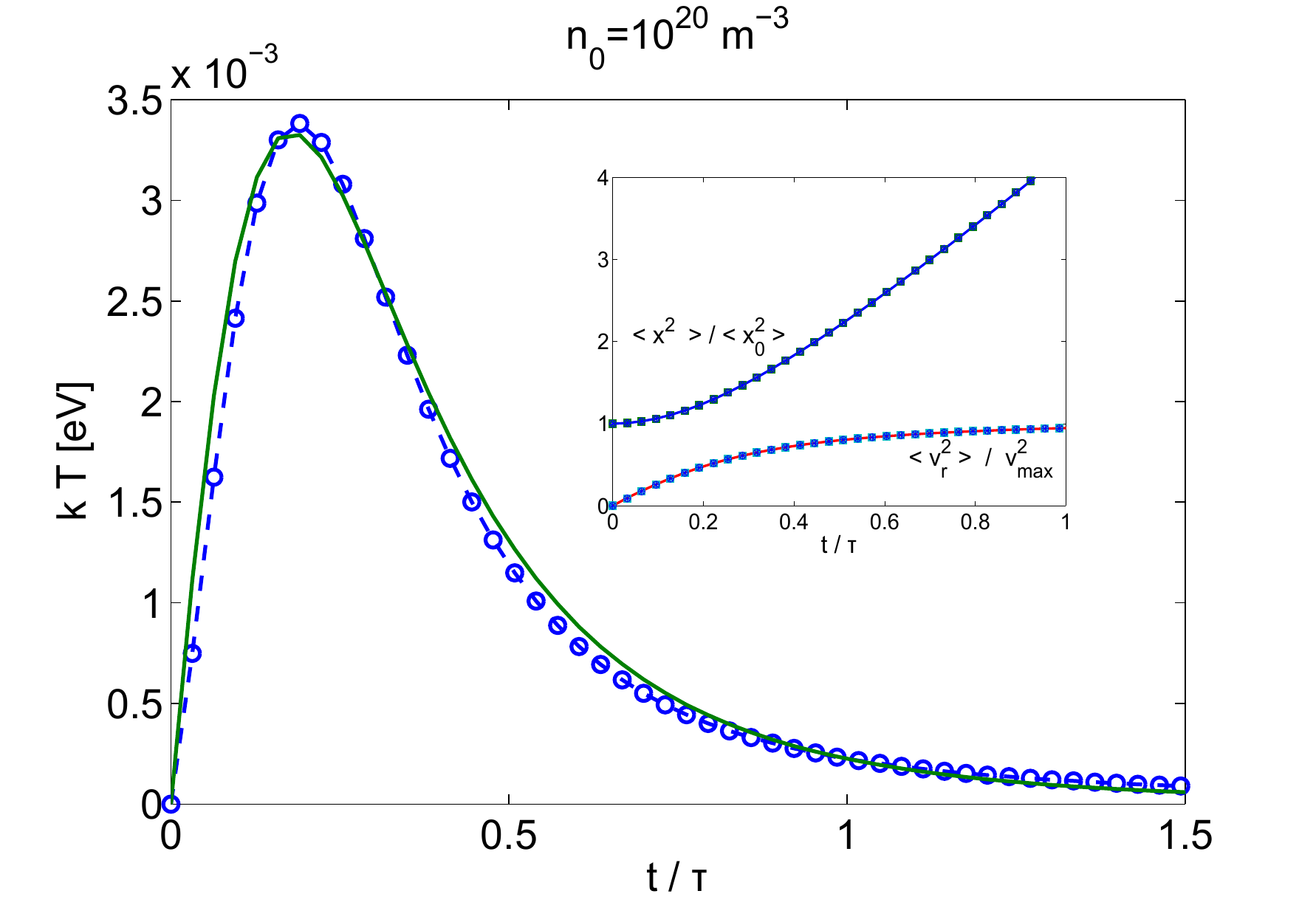}
\caption{\label{both} RMS velocity spread of a spherical bunch undergoing coulomb expansion, treecode data (circles), and prediction of \ref{mytheory}. The inset shows the overall expansion of the bunch in normalized units, both analytic prediction (solid line), given by Gauss's Law, and treecode data (dots). Uniform bunches remain uniform, and the normalized bunch size $R(\tau)/R_0$ is independent of density. }
\end{figure}

Simulations were performed to determine the temperature as a function of time for a bunch undergoing full Coulomb explosion. Multiple densities were considered, however the previous scaling with $n_0^{1/3}$ was seen. Thus, we present only results for $n_0=10^{20}$ \m in figure \ref{both}. The total overall expansion and velocity growth is plotted in the inset, with the analytic prediction, showing excellent agreement. The temperature shows an initial increase due to DIH at a time of $t_{\text{max}}\approx\omega_p^{-1}$, to a value of $kT (t=t_{\text{max}})=e^2/4\pi \epsilon_0 a(t_{\text{max}})$, or a value $a(t_{\text{max}})/a_0 \approx 1.6$ times smaller than for a fully contained beam. The bunch continues to expand, attempting to equilibrate to $\Gamma_{eq}=2.23$, and the temperature continues to decrease. However, at longer times, the temperature falls as $a^{-2}$, as in an uncoupled system. Thus, there must be a time at which correlation ceases to grow with decreasing temperature. Such behavior has been noted in the adiabatic expansion of UNPs due to kinetic pressure \cite{decouple}.

Thus, to model the temperature as a function of time, we can write the temperature as a product of the correlated growth, and the decoupled expansion after correlation ceases. Since the growth of the radius is small during the time of DIH, we may presume $T=T_{c}(t)\frac{a_0^2}{a(t)^2}$, where $T_c$ is the correlated temperature dependence, and the term $\frac{a_0^2}{a(t)^2}$ describes the expansion of independent particles. The rate of DIH should be proportional to the how far the system is from the equilibrium temperature at the current density, and so:
\begin{equation}
\frac{dT_c}{dt}=\frac{T_{c,eq}-T_c(t)}{\tau_0}=\frac{1}{\tau_0}\left( \frac{e^2}{4 \pi \epsilon_0 a(t) \Gamma_{eq}} -T_c(t) \right)
\label{mytheory}
\end{equation}

\begin{figure}[ht]
\includegraphics[width=1.0\linewidth, clip]{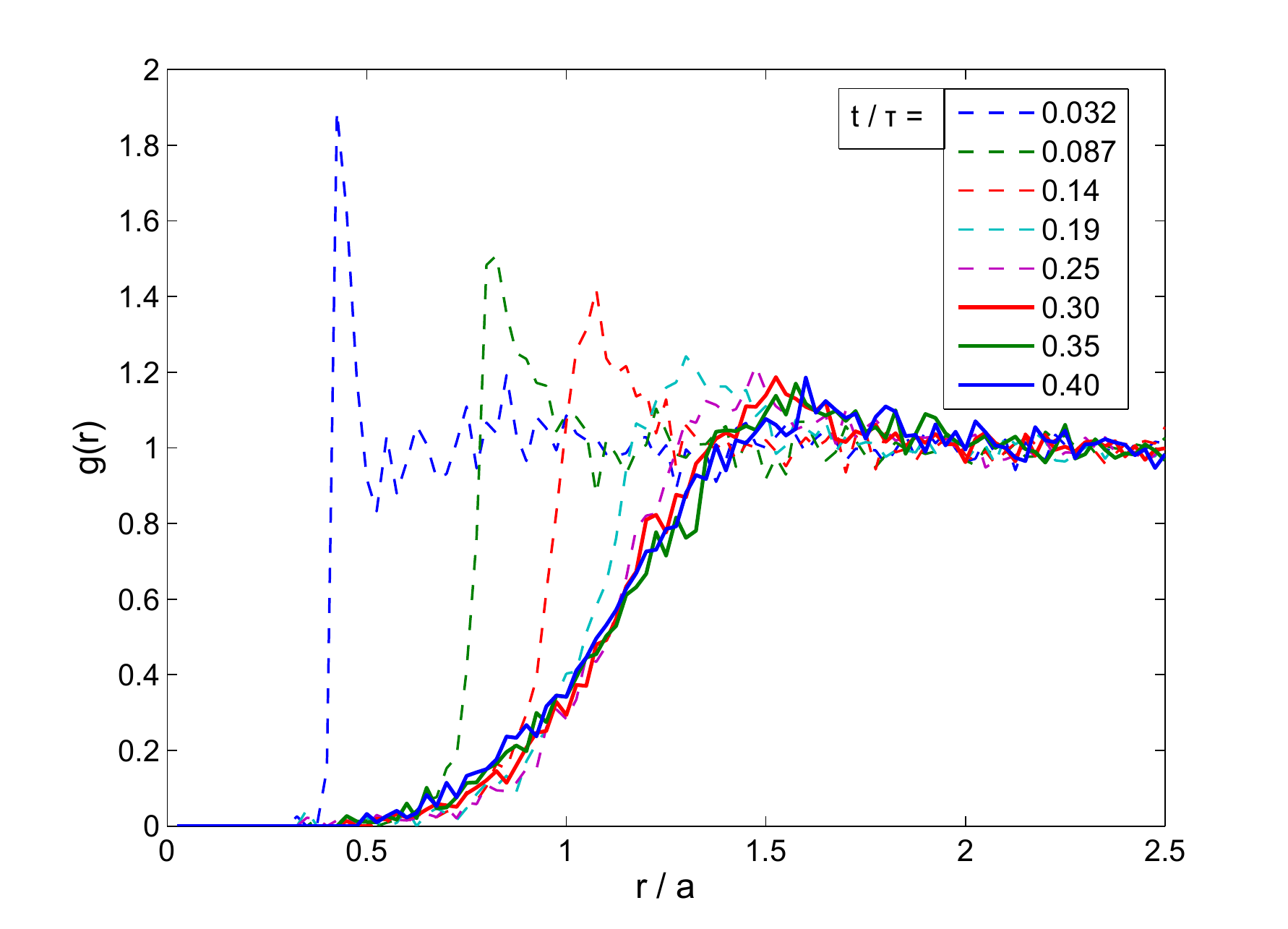}
\caption{\label{gofr} Pair correlation function g(r), computed at multiple simulation times for a bunch undergoing coulomb explosion. Both the initial ``shock" profile and the freezing out of correlations despite increasing $\Gamma$ is visible.}
\end{figure}

The only free parameter of this model is decoupling time $\tau_0$, or the time when the correlation ceases to increase. Integrating this numerically, and fitting to the temperature data, we find $\tau_0=0.325\tau$, which is also plotted in figure \ref{both}. This time is non-negligibly greater than $t_{\text{max}}$, and suggests that the bunch is cooled not only via decoupled expansion, but coupled expansion as well just after $t_{\text{max}}$.  We can verify this prediction by looking at the pair correlation as a function of time. This is computed in figure \ref{gofr}, via averaging over $5\times10^{3}$ particles randomly selected from the distribution. The development of the ``Coulomb hole" and an accompanying shock profile due to violent repulsion is clearly visible, as has also been seen in UNP simulations \cite{blast}. Indeed the correlation ceases to change after a time of approximately $\tau_0=0.3\tau$. Partial confinement effected by linear focusing in the source would yield an altered $a(t)$, and thus a longer $\tau_0$, whereas acceleration that lengthens the bunch significantly near the DIH heating timescale would yield a shorter $\tau_0$, but in either case the DIH heating timescale should remain unchanged. 

The exploding beam case justifies \textit{post-factum} the validity of the temperature limit of \ref{mymurillo} for practical photoemission, though beams in many applications do not have constant volume. In the exploding case, near $\tau_0$ the temperature scaling shifts from $a(t)^{-1}$ as given in Eq. 2, to the familiar scaling of $\sigma_v^2\propto a(t)^{-2}$, as given by emittance/brightness conservation. A more detailed calculation of DIH for practical cases in which the bunch shape is both time and space dependent should involve averaging over a temperature distribution given by \ref{mymurillo} and \ref{mytheory} using the local density as determined by space charge tracking.

\section{Conclusion}
In summary, we have characterized the fundamental temperature limit of photoemission from the disorder induced heating of electrons due to poorly screened Coulomb interactions at all length scales, where the analytic two particle models fail. We have shown that the tabulated thermodynamic quantities of one component plasmas are sufficient to explain both fully contained and Coulomb exploding instances of DIH, and have verified two simple relations that describe the temperature evolution.  Furthermore, many interesting effects of DIH in UNP, such as temperature oscillation and correlation decoupling,  should also be present in such cold beams, yielding the possibility of rich interdisciplinary study. 

Practically, for next-generation ultracold dense electron sources we have shown that the temperature of photoemission, and thus the maximum beam brightness, cannot be arbitrarily improved. Furthermore, given the rapid progress of photocathode temperature reduction, we anticipate such a limit to be approached in the next generation of high brightness electron sources producing intense beams.  

\ack

This work was supported by the National Science Foundation grant DGE-0707428, and by the Department of Energy grant DE-SC0003965. This work was also supported by the Director, Office of Science, Office of Basic Energy Sciences of the U. S. Department of Energy, under Contract Nos. DE-AC02-05CH11231, KC0407-LSJNT-I0013, and DE-SC0005713.
\\

\bibliographystyle{unsrt}
\bibliography{DIH}{}

\begin{thebibliography}{10}

\bibitem{ivan}
Ivan~V. Bazarov, Bruce~M. Dunham, and Charles~K. Sinclair.
\newblock Maximum achievable beam brightness from photoinjectors.
\newblock {\em Phys. Rev. Lett.}, 102:104801, Mar 2009.

\bibitem{cryo}
S.~Pastuszka, D.~Kratzmann, D.~Schwalm, A.~Wolf, and A.~S. Terekhov.
\newblock Transverse energy spread of photoelectrons emitted from gaas
  photocathodes with negative electron affinity.
\newblock {\em Applied Physics Letters}, 71(20):2967--2969, 1997.

\bibitem{narrowcone}
Zhi Liu, Yun Sun, P.~Pianetta, and R.~F.~W. Pease.
\newblock Narrow cone emission from negative electron affinity photocathodes.
\newblock volume~23, pages 2758--2762. AVS, 2005.

\bibitem{MOT}
G.~Taban, M.~P. Reijnders, B.~Fleskens, S.~B. van~der Geer, O.~J. Luiten, and
  E.~J.~D. Vredenbregt.
\newblock Ultracold electron source for single-shot diffraction studies.
\newblock {\em EPL (Europhysics Letters)}, 91(4):46004, 2010.

\bibitem{jansen}
G.H. Jansen.
\newblock {\em Coulomb interactions in particle beams}.
\newblock Advances in electronics and electron physics: Supplement. Academic
  Press, 1990.

\bibitem{reiser}
M.~Reiser.
\newblock {\em Theory and Design of Charged Particle Beams}.
\newblock Wiley Series in Beam Physics and Accelerator Technology. John Wiley
  \& Sons, 2008.

\bibitem{massey}
G.~A. Massey, M.~D. Jones, and B.~P. Plummer.
\newblock Space-charge aberrations in the photoelectron microscope.
\newblock {\em Journal of Applied Physics}, 52(6):3780--3786, 1981.

\bibitem{teller}
S.~G. Brush, H.~L. Sahlin, and E.~Teller.
\newblock Monte carlo study of a one-component plasma. i.
\newblock {\em The Journal of Chemical Physics}, 45(6):2102--2118, 1966.

\bibitem{referencefarm}
M~Lyon and S~D Bergeson.
\newblock The influence of electron screening on disorder-induced heating.
\newblock {\em Journal of Physics B: Atomic, Molecular and Optical Physics},
  44(18):184014, 2011.

\bibitem{verifymurillo}
Y.~C. Chen, C.~E. Simien, S.~Laha, P.~Gupta, Y.~N. Martinez, P.~G. Mickelson,
  S.~B. Nagel, and T.~C. Killian.
\newblock Electron screening and kinetic-energy oscillations in a strongly
  coupled plasma.
\newblock {\em Phys. Rev. Lett.}, 93:265003, Dec 2004.

\bibitem{Murillo}
M.~S. Murillo.
\newblock Using fermi statistics to create strongly coupled ion plasmas in atom
  traps.
\newblock {\em Phys. Rev. Lett.}, 87:115003, Aug 2001.

\bibitem{review}
Daniel H.~E. Dubin and T.~M. O'Neil.
\newblock Trapped nonneutral plasmas, liquids, and crystals (the thermal
  equilibrium states).
\newblock {\em Rev. Mod. Phys.}, 71:87--172, Jan 1999.

\bibitem{plasmatreecode}
Byoungseon Jeon, Joel~D. Kress, Lee~A. Collins, and Niels Grønbech-Jensen.
\newblock Parallel tree code for two-component ultracold plasma analysis.
\newblock {\em Computer Physics Communications}, 178(4):272 -- 279, 2008.

\bibitem{gpt}
Pulsar physics, general particle tracer v. 3.10.

\bibitem{barnesonly}
Joshua Barnes.
\newblock Treecode guide, February 2001.

\bibitem{barneshut}
Josh Barnes and Piet Hut.
\newblock {A hierarchical O(N log N) force-calculation algorithm}.
\newblock {\em Nature}, 324(6096):446--449, December 1986.

\bibitem{decouple}
T.~Pohl, T.~Pattard, and J.~M. Rost.
\newblock Relaxation to nonequilibrium in expanding ultracold neutral plasmas.
\newblock {\em Phys. Rev. Lett.}, 94:205003, May 2005.

\bibitem{blast}
Michael~S. Murillo.
\newblock Ultrafast dynamics of strongly coupled plasmas.
\newblock {\em Phys. Rev. Lett.}, 96:165001, Apr 2006.

\end{thebibliography}

\end{document}